# Re–compression Based JPEG Tamper Detection and Localization Using Deep Neural Network, Eliminating Compression Factor Dependency

Jamimamul Bakas[✉], Praneta Rawat, Kalyan Kokkalla, and Ruchira Naskar

Department of Computer Science and Engineering, National Institute of Technology, Rourkela 769008, India
{516cs6008,naskarr}@nitrkl.ac.in, pranetarawat1994@gmail.com, kalyan.kokkalla@gmail.com

**Abstract.** In this work, we deal with the problem of re–compression based image forgery detection, where some regions of an image are modified illegitimately, hence giving rise to presence of dual compression characteristics within a single image. There have been some significant researches in this direction, in the last decade. However, almost all existing techniques fail to detect this form of forgery, when the first compression factor is greater than the second. We address this problem in re–compression based forgery detection, here Recently, Machine Learning techniques have started gaining a lot of importance in the domain of digital image forensics. In this work, we propose a Convolution Neural Network based deep learning architecture, which is capable of detecting the presence of re–compression based forgery in JPEG images. The proposed architecture works equally efficiently, even in cases where the first compression ratio is greater than the second. In this work, we also aim to localize the regions of image manipulation based on re–compression features, using the trained neural network. Our experimental results prove that the proposed method outperforms the state–of–the–art, with respect to forgery detection and localization accuracy.

Digital forensics · Double compression · Image forgery
Joint photographic experts group (JPEG)
Re–compression based forgery

## 1 Introduction

In today's world, majority of day–to–day communication relies on exchange of digital data. Hence, assuring the trustworthiness of their contents is crucial. Images play a very important role in present–day digital world, where they form the primary means of communications, as well as the major sources of evidence towards any event, in legal, media and broadcast industries. Due to the present wide availability of low–cost image processing tools and software, digital images have become highly vulnerable to illegitimate modification attacks. Due to the availability of such tools, *doctored photographs* have become wide–spread, which challenge the forensic analysts and research community greatly. The threat to the integrity and authenticity of digital images, has been further increased by the fact that most image manipulations are indiscernible to human eyes. From the past decade, the field of digital forensics has emerged to protect and restore the integrity and authenticity of digital data. Digital Forensics is the branch of science that deals with the investigation of doctored material found in digital devices related to computer crime. Traditional techniques, such as *Digital Watermarking* and *Digital Signature*, have been very widely adopted till date, for multimedia security and protection. However, a major drawback of these approaches is the requirement of data pre–processing. That is, they involve some precautionary measures, always. This makes such techniques limited to the specially equipped cameras, with specific embedded software and hardware chips. Such security measures are termed *active* techniques [1]. On the contrary, forensic techniques are *passive* (also known as blind) [1]. *Passive techniques* require no a–priori information processing or computation, and are completely based on post–processing of data. This forensics techniques are based on the assumption that digital forgeries alter the underlying statistics of an image, and leave behind *traces*, which may be intelligently exploited in the future to detect the forgeries and their sources.

Joint Photographic Expert Group (JPEG) [2] is the most widely used format for an image data storage, due to its best compression features and optimal space requirement. Substantial research has been carried out in the domain of JPEG forgery detection in the recent years [3–6]. The JPEG attack model considered by the researchers is as follows. A JPEG image is shown in Fig. 1(a). Let $QF_1$ denotes the initial quality factor, at which this image was JPEG compressed. A region of the image, as shown in Fig. 1(b) has been extracted and re–compressed at a different ratio $QF_2$, such that $QF_2 \neq QF_1$. The extracted region is put back into the original image, (at the same location), to produce the tampered image, shown in Fig. 1(c). The resultant image is nothing but another JPEG, consisting of two differently compressed regions, one doubly compressed with subsequent quality factors $QF_1, QF_2$, and the rest of the image singly compressed at $QF_1$, as shown in Fig. 1(c). It is evident from Fig. 1(c), that the tampered region having a (double) compression ratio, different from the rest of the image, is perceptually indistinguishable.

In this paper, we focus on the detection and localization of double compression based JPEG modification attack, modelled as above. In this work, we model the given challenge as a two–way classification problem. However, conventional machine learning based classifiers are solely based on *feature identification and extraction*. Such conventional classification techniques prove to be inefficient for problems, in which the features are not identified or well–known, or their extraction is difficult. To address this issue, in this work, we develop a *Convolution*

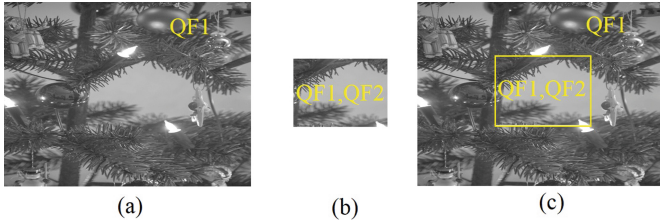

(a)                    (b)                    (c)

**Fig. 1.** JPEG attack on image: (a) Original $512 \times 512$ image; (b) Central region, re–saved at a different compression ratio; (c) Forged image with differently compressed regions

*Neural Network* (CNN) based *deep learning* network architecture, which would assist in automated feature engineering in classification task. Our first aim is to perform a two–way classification, between (A) unforged (single compressed) JPEG images, and (B) forged (double compressed) JPEG images. Our second aim is localization of forged region(s) in a JPEG image. We achieve this by performing a JPEG block–wise CNN classification, applied to our test images. The performance of the proposed forgery localization method has been improved, by considering vertical and horizontal strides of magnitude, as low as eight pixels. This helped us achieve forgery localization units upto $32 \times 32$ pixels, hence improving the detection accuracy as compared to the state-of-the-art. This is evident from our experimental results. Additionally, the proposed method successfully addresses those cases of re–compression based JPEG forgeries where the first compression factor is greater than the second ($QF1 > QF2$), unlike other state–of–the–art techniques such as [4,7,8]. Our experimental results prove this.

Rest of the paper is organized as follows. In Sect. 2, we provide an overview of the related background. In Sect. 3, we present the proposed CNN model for JPEG forgery detection and localization, along with the details of its attributes and architecture. In Sect. 4, we present our experimental results and related discussion. Finally, we conclude the paper with future research directions in Sect. 5.

## 2   Related Work

In this section, we review the existing literature on JPEG image forgery detection and localization. In this paper, we adopt a blind digital forensic approach to address the above problem, and here we present an overview of the related researches in this domain.

A number of significant researches towards double compression based JPEG forgery detection, are based on *Benford's Law* [9–11]. *Benford's law* or *first–digit law*, gives a frequency distribution prediction of the most significant digits in real–life numeric data sets. We focus on detection of image tampering in this paper by checking the DQ effect of the double quantized DCT coefficients. The DQ effect is the exhibition of periodic peaks and valleys in the distributions of the DCT coefficients. Related researches based on exploiting DCT coefficient are listed below.

In [9], the authors had investigated and analyzed the frequency distribution or histogram of DCT coefficients of JPEG images, for re–compression based JPEG forgery detection. Double quantization introduces specific artifacts into a JPEG, which is evident from its DCT coefficients histogram. These artifacts have been exploited in [9], for JPEG forgery detection. In [11], the authors proposed a JPEG forgery detection model, based on statistical analysis of the DCT quantization coefficients distribution, using generalized Benford Distribution Law. Among the other significant works based on Discrete Cosine Transform (DCT) coefficients distribution analysis utilizing generalized Benford's Law, for JPEG double compression detection, are [7,12–14]. The first significant attempt to localize tampered regions in JPEG images, was made by [15], using DCT of overlapping blocks and their lexicographical representations. [9] proposed a *block matching algorithm* to strike a balance between performance and complexity of such methods. Here, the authors adopted Principal Component Analysis (PCA) for image block representation.

Recently, neural networks have started gaining huge popularity in image forgery detection and classification tasks, due to spontaneous feature learning capabilities of such networks, which help to maximize classification accuracy. In [16], Gopi et al. utilized Artificial Neural Network (ANN) based classification and auto regressive image coefficients to generate feature vectors. The authors trained the network with 300 manually tampered training images, and tested the model with a different test set of 300 images. They achieved a forgery detection success rate of 77.67%. In [17], Bayar et al. developed a Convolution Neural Network (CNN) architecture which automatically learns image manipulation features, directly form the training data. In [5], Cozzolino et al. proposed a JPEG forgery detection scheme, which extracts image local residual features by means of a CNN. They fine–tuned the network with the labeled data and performed classification based on the extracted features.

The authors in [6] utilized a CNN to automatically learn hierarchical pattern representations from RGB color images. The pre–trained CNN is used as a patch descriptor to extract dense features from the test images, and to convert it to a more abstract form.

In [18], the authors address the problem of aligned and non–aligned forgery detection in JPEG images. They provided three solutions. The first involving handcrafted features extracted from JPEG, and a feature fusion technique is then adopted to obtain the final discriminative features for SVM classification. In the rest two, the CNN is directly trained with JPEG and with denoised images. CNN based on hand-crafted features allows us to achieve better accuracy than the other two methods, and performs efficiently when the second quality factor is greater than the first.

# 3    Proposed Deep Learning Model for Double Compression Based JPEG Forgery Detection and Localization

In this section, we present the proposed Convolution Neural Network based model for double–compression based JPEG forgery detection, as well as localization. The proposed forgery detection method consists of an initial JPEG pre–processing phase, followed by CNN learning. The trained CNN is later used for forgery localization in tampered JPEG images. For training of the proposed model, we use the following two datasets: (A) A set of images collected from the [19] uncompressed image database, which are subsequently compressed using JPEG with quality factor $QF_1$ (say); this serves as our authentic singly compressed image dataset ($S_{SC}$). (B) A second set of images which are generated by re–compressing the images in $S_{SC}$, this time by JPEG quality factor $QF_2$. This set forms our second training dataset of doubly compressed JPEG images, with quality factor ($QF_1, QF_2$); we name this dataset as $S_{DC}$.

In the pre–processing phase of the proposed method, we divide all images in $S_{SC}$ and $S_{DC}$, into $32 \times 32$ overlapping blocks, with a stride of 8 pixels. From each such block, a $19 \times 7$ dimensional feature vector (based on DCT frequency histogram [9]) is obtained in this phase; hence generating two sets of features: $F_{SC}$ and $F_{DC}$, from datasets $S_{SC}$ and $S_{DC}$, respectively. We label the samples belonging to $S_{SC}$ with 0, and those belonging to $S_{DC}$ with 1, in the pre–processing phase.

The next phase of the proposed method is the CNN learning phase. In this phase, we train the proposed CNN model with $F_{SC}$ and $F_{DC}$, i.e., the features obtained from singly compressed verses doubly compressed training images. The above features efficiently distinguish between single compressed and double compressed JPEG images, as evident from our experimental results in Sect. 4.

By specifying the features in the pre–processing step, we reduce the burden of feature engineering on the proposed CNN, so that it can focus more on dealing with tampered region localization. This considerably helps in complexity optimization. For forgery localization, the proposed CNN learns the hidden feature representations of artifacts caused due to tampering. The unit of forgery localization in the proposed method, is determined by the magnitude of block stride (used while division of an image into overlapping blocks, in the pre–processing phase), which is $8 \times 8$ in our work. This maximizes the forgery localization accuracy of the proposed method. Detailed experimental results are presented in Sect. 4.

Next, we describe the phases of the proposed method in detail, along with description of the proposed CNN architecture.

## 3.1    Pre–processing and Feature Extraction

The major task in JPEG pre–processing phase of the proposed method is extraction of block–wise features, depending on which we train the proposed CNN model. As stated previously, we divide the image into overlapping $W \times W = 32 \times 32$ blocks, with a stride of $S = 8$. For an $M \times N$ image, we obtain a total of $(\lceil \frac{M-W}{S} \rceil + 1) \times (\lceil \frac{N-W}{S} \rceil + 1)$ blocks of size $32 \times 32$ pixels.

For CNN learning, we use distributions of 19 DCT coefficients of each $32 \times 32$ image block, starting from second to the twentieth coefficients, in zigzag order, as feature vectors. Since each $32 \times 32$ image block consists of 16 $8 \times 8$ DCT blocks, we have 16 different values of each component (component 2 to component 20). For the $i$–th component, we find the block where it assumes the highest value as compared to the rest 15 blocks. We consider this block, and its six neighbors: position–wise its three immediate predecessor and three immediate successor blocks, for feature extraction. That is, if the block containing the highest value for component $i$ is indexed 0, we consider blocks indexed $[-3, -2, -1, 0, 1, 2, 3]$, for feature vector generation. This generates a $19 \times 7$ feature vector for each

**Input** : Input image $I$ of dimension $M \times N$
**Output**: Feature matrix $F$.

Initialize $W \leftarrow blocksize$;
Initialize $S \leftarrow stride$;
Initialize $n\_hor\_blocks \leftarrow (\lceil \frac{M-W}{S} \rceil + 1)$;
Initialize $n\_ver\_blocks \leftarrow (\lceil \frac{N-W}{S} \rceil + 1)$;
Initialize $F\_row \leftarrow 1$ // Row index to feature matrix $F$, every row of which stores
    $19 \times 7$ features extracted from each $W \times W$ image block
**for** $i$ from 1 to $(8 \times n\_hor\_blocks - 7)$ in steps of $S$ **do**
    **for** $j$ from 1 to $(8 \times n\_ver\_blocks - 7)$ in steps of $S$ **do**
        /* Processing one $32 \times 32$ image block                          */
        $block \leftarrow I(i : i + W - 1, j : j + W - 1)$;
        Initialize $block\_cnt \leftarrow 0$ // Counter for DCT blocks
        **for** $p$ from 1 to 4 **do**
            **for** $q$ from 1 to 4 **do**
                /* Feature extraction from $8 \times 8$ DCT blocks         */
                $sub\_block \leftarrow block(8p - 7 : 8p, 8q - 7 : 8q)$;
                $dct\_sub\_block \leftarrow DCT(sub\_block)$;
                $block\_cnt \leftarrow block\_cnt + 1$;
                $f\_vector(block\_cnt, 1 : 19) \leftarrow dct\_sub\_block(2 : 20)$ // store nineteen
                    coefficients for each DCT block
            **end**
        **end**
        /* Generating feature matrix for each $32 \times 32$ image block        */
        **for** $c$ from 1 to 19 **do**
            // Computing max value for coefficient $c$, finding its position, and six
                neighboring DCT blocks
            Initialize $max \leftarrow f\_vector(1, c)$ // Initializing the maximum value with
                value at DCT block 1, for each coefficient $c$
            Initialize $max\_pos = 1$ // To store block index containing maximum value of
                coefficient $c$
            **for** $block\_cnt$ from 1 to 16 **do**
                **if** $f\_vector(block\_cnt, c) > max$ **then**
                    $max = f\_vector(block\_cnt, c)$;
                    $max\_pos = block\_cnt$;
                **end**
            **end**
            $F(F\_row, 7c - 6 : 7c) = f\_vector(max\_pos - 3 : max\_pos + 3, c)^T$ ;
        **end**
        $F\_row \leftarrow F\_row + 1$;
    **end**
**end**

**Algorithm 1:** JPEG Pre–processing

$32 \times 32$ image block, in our work. This abstraction is carried out to reduce computational complexity, without losing any significant block information.

To present the DCT coefficient selection procedure more clearly to the readers, we present an example here, in Fig. 2, which shows a $32 \times 32$ image block, consisting of 16 $8 \times 8$ DCT blocks. In Fig. 2, we can see that the second coefficient assumes values $2.185e^{-16}$, $8.283e^{-16}$ etc. over the different DCT blocks, sequentially. The second coefficient assumes its highest value $9.409e^{-16}$, at the $(4, 1)$–th DCT block. Hence, to generate features corresponding to the second DCT coefficient of the given image block, we consider the $(4,1)$–th DCT block, along with its three preceding and three succeeding neighbours, that is, DCT blocks: $(3, 2)$, $(3, 3)$, $(3, 4)$, $(4, 1)$, $(4, 2)$, $(4, 3)$, $(4, 4)$. The 7–dimensional feature vector, corresponding to the second DCT coefficient of the given image block is: $[2.1852e^{-16}, 2.185e^{-16}, 2.185e^{-16}, 9.409e^{-16}, 4.968e^{-16}, 4.9688e^{-16}, 4.992e^{-16}]$. Similarly, we extract eighteen more 7–dimensional feature vectors from the rest of the coefficients, from third to twentieth; hence generating a $19 \times 7$ feature vector for each $32 \times 32$ image block, which is fed to the proposed CNN model, described next.

The set of feature vectors, thus obtained from images, belonging to sets $S_{SC}$ and $S_{DC}$, are denoted as $F_{SC}$ and $F_{DC}$ respectively. The complete pre–processing and feature extraction phase is presented in form of Algorithm 1.

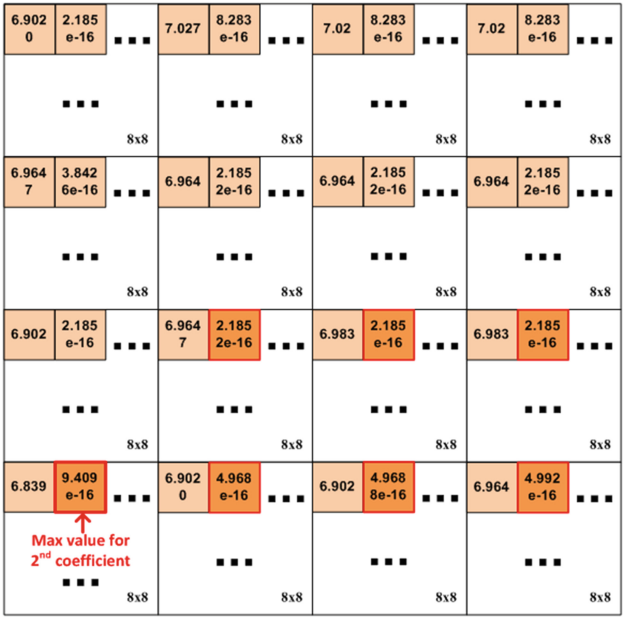

**Fig. 2.** Example: feature vector generation from DCT coefficients (*second* coefficient shown)

## 3.2  CNN Architecture

Convolution Neural Networks (CNN) form a variation of Multilayer Perceptrons (MLP), which consist of *neurons*, and *learnable biases* which are dependent on factors including local receptive fields, shared weights, spatial and temporal sub–sampling. CNNs consist of successive Convolution and sub–sampling layers, which are alternated, and finally connected to a Fully–connected layer. Convolution layers are responsible for performing a local feature average, and the sub–sampling layer, also called *Pooling* layer, is responsible for dimensionality reduction of the feature map. The Fully–connected layer implicitly consists of two more layers: *Dense layer* and *Logits layer*. The Dense layer performs classification based on features extracted by the previous convolution/pooling layers. Further, the Logits layer produces the raw prediction values. Each layer of a CNN receives input from the previous layer, multiplied by appropriate learnable weights, and are further added with biases.

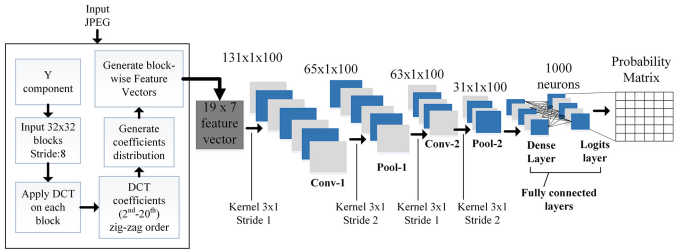

**Fig. 3.** Convolution Neural Network (CNN) architecture

As discussed in Sect. 3.1, we obtain features $F_{SC}$ and $F_{DC}$, from our single compressed and double compressed training images, respectively. Each of $F_{SC}$ and $F_{DC}$, is a matrix where each row consists of $19 \times 7$ features, and corresponds to one $32 \times 32$ single or double compressed JPEG block, respectively. Next, we shuffle the rows of matrices $F_{SC}$ and $F_{DC}$, and hence merge those into a single matrix $F_{shuffled}$. Shuffling data serves the purpose of reducing variance in highly correlated examples, and ensuring that the classification model generalizes well and overfits less.

According to Sect. 3.1, we obtain $(\lceil \frac{M-32}{8} \rceil + 1) \times (\lceil \frac{N-32}{8} \rceil + 1)$ $32 \times 32$ blocks from an $M \times N$ image. We have used training/test images of size $384 \times 512$ pixels in our work, each of which generated 2,745 blocks, according to the above formulation. Our training dataset consists of 480 single and 480 double compressed JPEG images. Therefore, each of $F_{SC}$ and $F_{DC}$ consist of $480 \times 2,745$ = 1,317,600 feature vectors, and $F_{shuffled}$ consists of $1,317,600 \times 2 = 2,635,200$ feature vectors. Summarily, we train the proposed model using $480 \times 2 \times 2,745$ = 2,635,200 image blocks.

In this work, we propose a 2D–Convolution Neural Network architecture as shown in Fig. 3. In the proposed architecture, we adopt a $3 \times 1$ kernel at each layer, and vary the stride magnitude according to the required feature dimensionality. The input to the first convolution layer, Conv–1, is a $19 \times 7$ pixel matrix. Here we take a stride of 1 pixel, and the number of filters used in this layer is 100. After the first convolution layer, the output obtained is of dimension $131 \times 1 \times 100$, which is fed to the next layer. Here, 100 represents that there are 100 channels, each holding the output from one filter.

Pool–1 layer receives its input from Conv–1, and uses a stride of size 2; the number of filters used is 100. In this layer, our objective is data dimensionality reduction (sub–sampling) Hence, the stride magnitude is increased here to minimize feature dimension. Output of Pool–1 is $65 \times 1$ dimensional.

In Conv–2, the input is of dimension $65 \times 1$, the stride size and number of filters being exactly same as those in Conv–1. The output of Conv–2 serves as the input to Pool–2, the dimension of which is $63 \times 1$. The size of kernel, stride and number of filters in Pool–1 layer, are same as those in Pool–2. The output of Pool–2 layer is $31 \times 1$ dimensional.

The final layer is a Fully Connected convolution layer, which consists of the Dense and Logits layers. In the Dense Layer, we use 1000 neurons and the output is fed to a two–way softmax connection. In [20], it has been proven that deep neural networks with ReLUs perform efficiently while training with large databases and faster than tanh and other learning functions. In our network, Rectified Linear Units (ReLUs), with an activation function $f(x) = max(0, x)$, are used for each connection.

The input dimension to this layer is $31 \times 1$. To improve the training accuracy of the proposed model, we applied *dropout* regularization to the Dense layer. According to this phenomenon, during the training process, randomly selected neurons are *dropped–out* or ignored. This constraints the learning of the network by reducing dependency between neurons, hence avoiding overfitting. The Logits layer performs the final classification, thus producing the probability of each individual block, of being single compressed or double compressed.

The loss function used in this network, is the *Softmax Cross–Entropy* function at the last layer, which is *back propagated* through the network. We use *Softmax Cross-entropy* here, since a 2–way classification has been performed in this work. To optimize the loss during training, a learning rate of 0.001 and the *Stochastic Gradient Descent* optimizer, have been used.

### 3.3   Localizing the Tampered Regions

Localization of tampered regions in JPEG images, is accomplished during the testing phase of the proposed model. The model is trained as described in Sect. 3.2, where each $32 \times 32$ image block is assigned its class label ('0' for single compressed, and '1' for double compressed). During testing too, we divide an image into $32 \times 32$ blocks, using a stride of 8 pixels, similar to the training phase pre–processing. Now, each block is tested using the *trained* CNN model, and the final outcome is block–wise prediction of JPEG forgery (the tampered regions are labelled '1', indicating that the region is double compressed according to our JPEG attack model discussed in Sect. 1, and the authentic regions are labelled '0').

Although the class prediction is performed by the proposed CNN model for each $32 \times 32$ image block, the unit of JPEG forgery localization here, is $8 \times 8$ pixels. The reason can be explained following Fig. 4. As evident from Fig. 4, after processing and testing the first $32 \times 32$ block (Fig. 4(a)), the stride moves right by 8 pixels, hence targeting the second $32 \times 32$ block (Fig. 4(b)). In this situation, after the stride movement towards right by 8 pixels is complete, the previous prediction for the first block, remains preserved only for the first (leftmost) $32 \times 8$ pixels. The remaining $32 \times 24$ pixels are newly tested and assigned a new class label, same as that of block 2, as they form a part of the second $32 \times 32$ block. Similarly, after traversal of one complete row, the stride performs vertical move by 8 pixels, as shown in Fig. 4(c). Hence, effectively, after stride movement of 8 pixels horizontally and vertically, we are left with the old block 1 prediction, only constrained to the top–left $8 \times 8$ pixels. This is evident from Fig. 4(d). This mechanism helps us obtain unit of forgery localization in the proposed model, as low as $8 \times 8$ image blocks.

Following similar movement/stride pattern, we process each (overlapping) $32 \times 32$ JPEG block sequentially, assign its class label using the trained CNN, and move on to the next block. For the last overhead blocks, we pad the image with sufficient number of *zero* rows and columns. This method makes the proposed JPEG forgery localization process considerably accurate, the unit of localization being merely $8 \times 8$ image blocks.

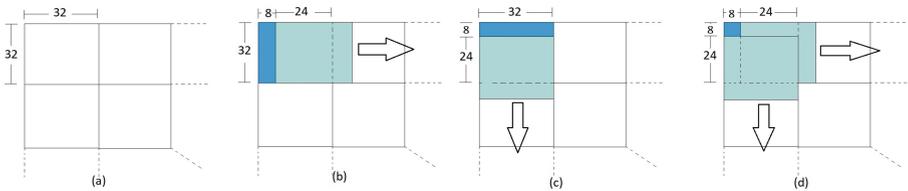

**Fig. 4.** Stride movement demonstration: (a) top–leftmost $32 \times 32$ image block, (b) Stride movement to the second block of the row, (c) First vertical stride movement to the second row, (d) Effective unit of forgery localization: top–leftmost $8 \times 8$ image block (in dark shade)

## 4  Experimental Results and Discussion

In this section, we first describe the dataset and the experimental set–up adopted by us, for performance evaluation of the proposed JPEG forgery detection and localization scheme. Then, we present our detailed experimental results. We compare the proposed method with recent state–of–the–art JPEG compression based forgery detection techniques, and present the relevant analysis results.

### 4.1  Dataset Generation and Experimental Set–Up

The JPEG pre–processing tasks in the proposed method, have been carried out using *MATLAB Image Processing Toolbox*. The proposed Convolution Neural Network has been implemented in *Tensorflow* parallel processing framework, in a *Python* environment.

In our experiments, we use 500 images collected from the UCID database [19]. All images provided in the UCID database, are in TIFF format, each of dimension $384 \times 512$ pixels. For our experiments, we first compress the TIFF images, with JPEG quality factor $QF_1 = 55, 65, 75, 85$ and 95. This way, we produce our single compressed image dataset $S_{SC}$, (described in Sect. 3). Next, the images in $S_{SC}$ are further re–compressed one more time, with quality factor $QF_2 = 55, 65, \cdots 95$; this time to generate our double compressed image set $S_{DC}$ (described in Sect. 3).

As discussed in Sect. 3.1, the JPEG images undergo a preliminary pre–processing step, before being used for training the CNN.

Out of the 500 images used in our experiment, we used 480 images for training, which generated a total of 1,317,600 blocks for training.

According to the JPEG modification model described in Sect. 1, we tamper our test JPEG images as follows. Some region of an image, compressed with quality factor $QF_1$ initially, has been modified, and saved at a different quality factor $QF_2$, to bring about re–compression based JPEG forgery. In particular, for our experiments, we have manually forged the test JPEG images, by replacing some region of a test image, initially compressed with quality factor $QF_1$, by the corresponding region, extracted from the same image, re–compressed at quality factor $QF_2$. We have varied the size of forgery as 10%, 30% and 50% of the actual images.

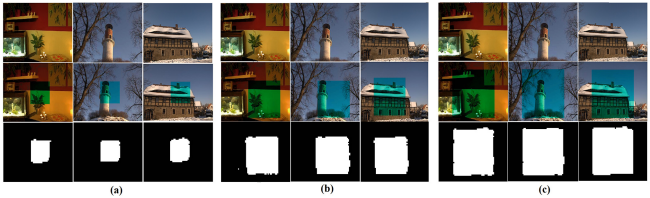

**Fig. 5.** Forgery detection and localization by the proposed method. Forgery sizes: (a) 10% (b) 30% (c) 50%. (*Top*) Authentic images. (*Middle*) Tampered images: tampered regions highlighted. (*Bottom*) Detected and Localized Forgeries

Forgery detection and localization results of the proposed method have been presented in Fig. 5, for three different forgery sizes.

### 4.2  Performance Evaluation Metrics

We model the problem of JPEG re–compression based forgery detection and localization, as a two–way classification problem, where we predict block–wise forgery. To evaluate the classification efficiency of the proposed method, we adopt a set of three metrics, viz. *Accuracy*, *F–measure*, *Success Rate*. We compare the proposed forgery detection method with con To evaluate the performance of the proposed forgery localization method, we use the following metric: of Forgery Localization, introduced by the authors in [4].

Accuracy of the proposed classification model can be defined as follows:

$$Accuracy = \frac{|TP| + |TN|}{|TP| + |TN| + |FP| + |FN|} \tag{1}$$

where $TP$, $TN$, $FP$ and $FN$ represent the sets of True Positive, True Negative, False Positive and False Negative samples, respectively.

The parameter *F–measure*, related to performance of a classification model, is also defined based on $TP$, $TN$, $FP$ and $FN$, as follows:

$$F - measure = \frac{2 \times Precision \times Recall}{Precision + Recall} \tag{2}$$

where,

$$Precision = \frac{|TP|}{|TP| + |FP|}, \ Recall = \frac{|TP|}{|TP| + |FN|} \tag{3}$$

In this paper, we report the F–measure averaged over $N = 20$ test images. Specifically, the reported F–measure is computed as:

$$F - measure = \frac{\sum\limits_{i=1}^{N} F - measure(i)}{N} \tag{4}$$

where *F-measure*$(i)$ gives the test results for the $i$–th image.

To evaluate the forgery localization efficiency of the proposed method, we follow the parameterization adopted by the authors in [4]. Here, a threshold $T_h$ is chosen, which determines that tampered regions in an image are correctly localized, iff $F - measure \geq T_h$. Similar to [4], we set $T = 2/3$ in this work. So, the third evaluation parameter used in this work, *Success Rate of Localization* is defined as follows:

$$Success \ Rate = \frac{\sum\limits_{i=1}^{N} \delta_{F-measure(i) \geq T_h}}{N} \tag{5}$$

where $N$ is the number of test images, and $\delta_{F-measure(i) \geq T_h}$ for every $i$–th image is computed as:

$$\delta_{F-measure(i) \geq T_h} = \begin{cases} 1 & \text{if } F - measure(i) \geq T_h, \\ 0 & \text{if others.} \end{cases}$$

**Table 1.** Performance evaluation and comparison for 10% Forgery: accuracy, F-measure and Success Rate of Localization results.

| QF1 | QF2 | | | 55 | 65 | 75 | 85 | 95 |
|-----|-----|---|---|-----|-----|-----|-----|-----|
| 55 | Proposed | Accuracy | | **56.683** | **90.833** | 93.073 | 93.073 | **95.473** |
| | | F−measure | | **29.563** | **73.473** | **82.173** | **86.393** | 90.963 |
| | | Success Rate | | 0 | **100** | **100** | **100** | **100** |
| | Wang et al. [4] | Accuracy | | - | 88.37 | 93.66 | 95.57 | 93.59 |
| | | F−measure | | - | 45.87 | 72.7 | 81.17 | 83.85 |
| | | Success Rate | | - | 39.01 | 70.63 | 80.19 | 83.56 |
| | Bianchi et al. [7] | Accuracy | | - | 90.02 | 79.53 | 86.25 | 74.82 |
| | | F−measure | | - | 67.33 | 54.17 | 65.43 | 50.16 |
| | | Success Rate | | - | 60.09 | 23.54 | 50.97 | 20.4 |
| | Lin et al. [8] | Accuracy | | - | 87.91 | 88.21 | 89.04 | 94.31 |
| | | F−measure | | - | 1.65 | 1.95 | 3.96 | 68.84 |
| | | Success Rate | | - | 0 | 0 | 2.84 | 70.1 |
| 65 | Proposed | Accuracy | | 65.353 | **57.643** | 83.253 | 85.193 | **96.593** |
| | | F−measure | | **44.313** | **36.333** | **75.083** | 77.383 | **89.823** |
| | | Success Rate | | 5 | 0 | **90** | **100** | **100** |
| | Wang et al. [4] | Accuracy | | 35.39 | - | 90.06 | 95.14 | 93.74 |
| | | F−measure | | 14.02 | - | 55.12 | 81.34 | 84.01 |
| | | Success Rate | | 0 | - | 49.33 | 80.57 | 83.87 |
| | Bianchi et al. [7] | Accuracy | | 82.87 | - | 86.1 | 86.08 | 65.84 |
| | | F−measure | | 41.02 | - | 64.16 | 66.62 | 40.31 |
| | | Success Rate | | 8.37 | - | 52.62 | 55.68 | 4.78 |
| | Lin et al. [8] | Accuracy | | 88.53 | - | 86.59 | 88.35 | 93.74 |
| | | F−measure | | 1.12 | - | 2.29 | 1.48 | 60.02 |
| | | Success Rate | | 0 | - | 0 | 0 | 61.41 |
| 75 | Proposed | Accuracy | | 74.493 | 70.843 | **58.023** | 86.583 | **96.573** |
| | | F−measure | | **60.793** | **59.363** | **31.123** | 76.653 | **85.673** |
| | | Success Rate | | **35** | **10** | 0 | **100** | **100** |
| | Wang et al. [4] | Accuracy | | 55.19 | 31.89 | - | 93.99 | 94.17 |
| | | F−measure | | 9.11 | 14.84 | - | 80.2 | 84.6 |
| | | Success Rate | | 0 | 0 | - | 79.45 | 83.48 |
| | Bianchi et al. [7] | Accuracy | | 88.57 | 88.54 | - | 64.36 | 80.22 |
| | | F−measure | | 7.24 | 24.65 | - | 43.48 | 59.52 |
| | | Success Rate | | 0 | 2.91 | - | 18.91 | 41.48 |
| | Lin et al. [8] | Accuracy | | 88.5 | 88.5 | - | 84.33 | 92.75 |
| | | F−measure | | 1.21 | 1.99 | - | 3.62 | 45.85 |
| | | Success Rate | | 0 | 0 | - | 0 | 47.68 |

<div align="right">(<em>continued</em>)</div>

**Table 1.** (*continued*)

| QF1 | QF2 | | 55 | 65 | 75 | 85 | 95 |
|---|---|---|---|---|---|---|---|
| 85 | Proposed | Accuracy | 80.833 | 76.563 | 71.233 | **57.693** | **95.693** |
| | | F–measure | **66.393** | **57.403** | **58.753** | 37.303 | **84.653** |
| | | Success Rate | **55** | **25** | **20** | 0 | **100** |
| | Wang et al. [4] | Accuracy | 43.2 | 24.31 | 21.94 | - | 93.02 |
| | | F–measure | 11.37 | 15.57 | 16.02 | - | 82.9 |
| | | Success Rate | 0 | 0 | 0 | - | 81.99 |
| | Bianchi et al. [7] | Accuracy | 76.42 | 26.28 | 87.78 | - | 44.91 |
| | | F–measure | 14.59 | 20.31 | 19.5 | - | 28.99 |
| | | Success Rate | 0 | 0 | 0.82 | - | 0.15 |
| | Lin et al. [8] | Accuracy | 88.19 | 86.65 | 84.46 | - | 89.48 |
| | | F–measure | 1.42 | 2.47 | 3.41 | - | 0.95 |
| | | Success Rate | 0 | 0 | 0 | - | 0.37 |
| 95 | Proposed | Accuracy | 83.573 | 81.433 | 76.903 | 75.713 | **62.323** |
| | | F–measure | **72.033** | **68.363** | **62.603** | **67.633** | **38.533** |
| | | Success Rate | **95** | **65** | **50** | **85** | 0 |
| | Wang et al. [4] | Accuracy | 37.77 | 25.43 | 12.75 | 34.5 | - |
| | | F–measure | 12.48 | 14.75 | 17.33 | 12.66 | - |
| | | Success Rate | 0 | 0 | 0 | 0 | - |
| | Bianchi et al. [7] | Accuracy | 73.19 | 76 | 65.81 | 61.5 | - |
| | | F–measure | 10.32 | 12.74 | 14.4 | 26.16 | - |
| | | Success Rate | 0 | 0 | 0 | 0 | - |
| | Lin et al. [8] | Accuracy | 88.04 | 86.43 | 83.26 | 75.75 | - |
| | | F–measure | 1.48 | 2.25 | 3.67 | 6.58 | - |
| | | Success Rate | 0 | 0 | 0 | 0 | - |

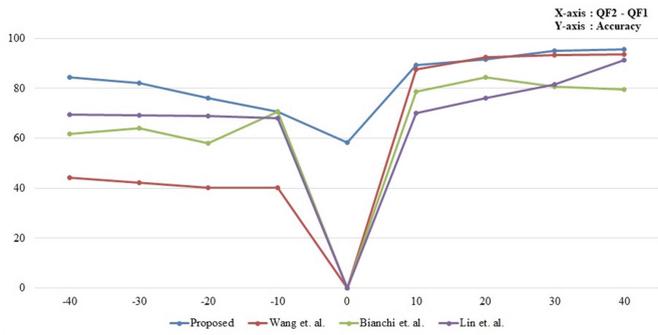

**Fig. 6.** Average accuracy for varying $QF_2 - QF_1$ values

**Table 2.** Performance evaluation and comparison for 30% Forgery: accuracy, F-measure and Success Rate of Localization results.

| QF1 | QF2 | | | 55 | 65 | 75 | 85 | 95 |
|---|---|---|---|---|---|---|---|---|
| 55 | Proposed | | Accuracy | **55.013** | **91.293** | **92.493** | **93.743** | **95.613** |
| | | | F−measure | **31.993** | 74.693 | 82.263 | 87.373 | 91.943 |
| | | | Success Rate | 0 | **100** | **100** | **100** | **100** |
| | Wang et al. [4] | | Accuracy | - | 89.9 | 91.03 | 93.65 | 94.1 |
| | | | F−measure | - | 62.88 | 87.02 | 90.3 | 92.59 |
| | | | Success Rate | - | 55.83 | 87.37 | 92.75 | 94.92 |
| | Bianchi et al. [7] | | Accuracy | - | 87.86 | 80.98 | 87.79 | 79.68 |
| | | | F−measure | - | 80.87 | 77.2 | 84.65 | 76.31 |
| | | | Success Rate | - | 82.44 | 81.46 | 91.7 | 74.96 |
| | Lin et al. [8] | | Accuracy | - | 69.62 | 69.95 | 73.6 | 91.46 |
| | | | F−measure | - | 5.56 | 7.49 | 18.92 | 81.96 |
| | | | Success Rate | - | 0.22 | 2.17 | 17.64 | 85.43 |
| 65 | Proposed | | Accuracy | 62.033 | **57.153** | 83.243 | 85.723 | **96.473** |
| | | | F−measure | 45.243 | **35.663** | 76.143 | 78.303 | 91.283 |
| | | | Success Rate | 5 | 0 | **95** | **100** | **100** |
| | Wang et al. [4] | | Accuracy | 43.12 | - | 85.02 | 93.62 | 94.06 |
| | | | F−measure | 36.05 | - | 73.08 | 90.61 | 92.36 |
| | | | Success Rate | 1.72 | - | 70.4 | 93.05 | 94.99 |
| | Bianchi et al. [7] | | Accuracy | 76.18 | - | 87.13 | 87.91 | 73.01 |
| | | | F−measure | 53.61 | - | 82.37 | 84.83 | 70.3 |
| | | | Success Rate | 33.63 | - | 84.86 | 89.31 | 59.72 |
| | Lin et al. [8] | | Accuracy | 69.9 | - | 69.13 | 69.71 | 90.75 |
| | | | F−measure | 2.67 | - | 8.07 | 10.71 | 78.79 |
| | | | Success Rate | 0.15 | - | 0.6 | 4.33 | 82.14 |
| 75 | Proposed | | Accuracy | **75.063** | 70.903 | **55.743** | 86.353 | **96.723** |
| | | | F−measure | **62.253** | **59.793** | **31.863** | 76.413 | 87.363 |
| | | | Success Rate | **45** | **20** | 0 | **100** | **100** |
| | Wang et al. [4] | | Accuracy | 48.57 | 40.03 | - | 92.11 | 94.57 |
| | | | F−measure | 26.36 | 36.79 | - | 89.06 | 93.09 |
| | | | Success Rate | 0.22 | 0.22 | - | 91.55 | 95.52 |
| | Bianchi et al. [7] | | Accuracy | 70.21 | 73.59 | - | 81 | 84.09 |
| | | | F−measure | 10.04 | 30.28 | - | 77.77 | 81.38 |
| | | | Success Rate | 0.3 | 4.11 | - | 76.31 | 83.93 |
| | Lin et al. [8] | | Accuracy | 69.94 | 69.49 | - | 67.67 | 88.53 |
| | | | F−measure | 2.7 | 4.69 | - | 12.11 | 69.66 |
| | | | Success Rate | 0.3 | 0.22 | - | 1.2 | 73.32 |

<div align="right">(<em>continued</em>)</div>

**Table 2.** (*continued*)

| QF1 | QF2 | | 55 | 65 | 75 | 85 | 95 |
|-----|-----|-----|-----|-----|-----|-----|-----|
| 85 | Proposed | Accuracy | **82.203** | **77.043** | 71.233 | **60.343** | **95.503** |
| | | F–measure | **69.143** | **58.413** | **60.323** | **37.763** | 84.783 |
| | | Success Rate | **75** | **25** | **25** | 0 | **100** |
| | Wang et al. [4] | Accuracy | 45.92 | 38.87 | 33.27 | - | 93.16 |
| | | F–measure | 29.66 | 38.22 | 43.37 | - | 91.3 |
| | | Success Rate | 0.15 | 0.22 | 0.22 | - | 94.62 |
| | Bianchi et al. [7] | Accuracy | 64.13 | 45.45 | 72.33 | - | 66.27 |
| | | F–measure | 23.25 | 50.65 | 26.05 | - | 65.66 |
| | | Success Rate | 0.37 | 0.15 | 1.42 | - | 45.44 |
| | Lin et al. [8] | Accuracy | 69.79 | 69.12 | 68.35 | - | 72.81 |
| | | F–measure | 3.18 | 5.23 | 7.6 | - | 15.22 |
| | | Success Rate | 0.15 | 0.22 | 0.45 | - | 13.15 |
| 95 | Proposed | Accuracy | **84.653** | **82.273** | **75.423** | **76.283** | **64.853** |
| | | F–measure | **74.133** | **69.933** | **64.163** | **68.033** | **37.753** |
| | | Success Rate | **100** | **85** | **65** | **85** | 0 |
| | Wang et al. [4] | Accuracy | 43.85 | 37.44 | 31.26 | 39.91 | - |
| | | F–measure | 31.84 | 38.29 | 44.67 | 35.09 | - |
| | | Success Rate | 0 | 0.15 | 0 | 0 | - |
| | Bianchi et al. [7] | Accuracy | 61.87 | 64.13 | 59.08 | 64.54 | - |
| | | F–measure | 19.18 | 22.09 | 28.55 | 50.8 | - |
| | | Success Rate | 0.22 | 0 | 0.37 | 1.79 | - |
| | Lin et al. [8] | Accuracy | 69.79 | 69.06 | 67.93 | 65.02 | - |
| | | F–measure | 3.3 | 4.95 | 8.39 | 15.29 | - |
| | | Success Rate | 0.22 | 0.22 | 0.15 | 0.52 | - |

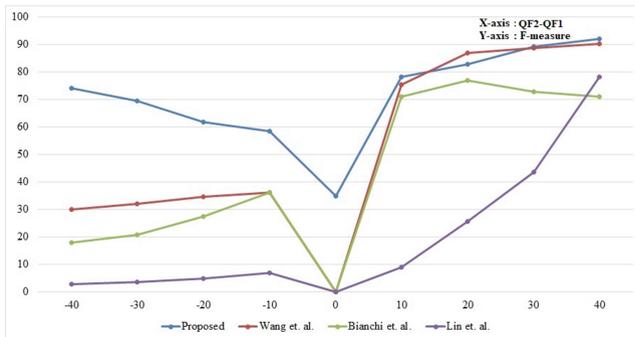

**Fig. 7.** Average F–measure for varying $QF_2 - QF_1$ values

**Table 3.** Performance evaluation and comparison for 50% Forgery: accuracy, F-measure and Success Rate of Localization results.

| QF1 | QF2 | | | 55 | 65 | 75 | 85 | 95 |
|-----|-----|-----|-----|-----|-----|-----|-----|-----|
| 55 | Proposed | | Accuracy | **56.603** | **90.493** | **93.053** | **94.243** | **96.073** |
| | | | F−measure | **32.073** | 76.323 | 83.833 | 88.103 | 93.413 |
| | | | Success Rate | 0 | **100** | **100** | **100** | **100** |
| | Wang et al. [4] | | Accuracy | - | 67.23 | 84.83 | 89.53 | 93.4 |
| | | | F−measure | - | 62.98 | 87.27 | 90.7 | 94.38 |
| | | | Success Rate | - | 72.27 | 96.94 | 97.91 | 99.48 |
| | Bianchi et al. [7] | | Accuracy | - | 83.66 | 83.33 | 89.85 | 84.32 |
| | | | F−measure | - | 82.58 | 86.02 | 91.2 | 86.96 |
| | | | Success Rate | - | 89.46 | 100 | 100 | 100 |
| | Lin et al. [8] | | Accuracy | - | 51.72 | 52.71 | 57.38 | 88.49 |
| | | | F−measure | - | 7.27 | 10.79 | 22.45 | 84.09 |
| | | | Success Rate | - | 3.74 | 8.15 | 23.54 | 89.69 |
| 65 | Proposed | | Accuracy | 63.233 | **57.783** | 84.353 | 85.233 | **96.923** |
| | | | F−measure | 46.343 | **34.413** | 76.113 | 79.373 | 92.633 |
| | | | Success Rate | 5 | 0 | **95** | **100** | **100** |
| | Wang et al. [4] | | Accuracy | 51.22 | - | 77.62 | 89.31 | 93.47 |
| | | | F−measure | 50.61 | - | 77.09 | 90.74 | 94.44 |
| | | | Success Rate | 70.4 | - | 85.35 | 98.51 | 99.55 |
| | Bianchi et al. [7] | | Accuracy | 63.92 | - | 85.38 | 89.71 | 81.49 |
| | | | F−measure | 48.87 | - | 85.84 | 91.1 | 84.86 |
| | | | Success Rate | 27.5 | - | 94.17 | 100 | 100 |
| | Lin et al. [8] | | Accuracy | 50.81 | - | 52.45 | 54.45 | 85.31 |
| | | | F−measure | 3.38 | - | 11.1 | 17.52 | 77.43 |
| | | | Success Rate | 1.12 | - | 6.13 | 15.84 | 82.51 |
| 75 | Proposed | | Accuracy | **75.523** | 71.113 | **54.813** | 86.443 | **97.063** |
| | | | F−measure | **64.213** | **62.283** | **31.803** | 76.933 | 88.763 |
| | | | Success Rate | **70** | 50 | 0 | **100** | **100** |
| | Wang et al. [4] | | Accuracy | 50.7 | 50.8 | - | 87.92 | 94.51 |
| | | | F−measure | 40.91 | 54.4 | - | 89.45 | 95.26 |
| | | | Success Rate | 54.56 | 75.34 | - | 97.83 | 99.4 |
| | Bianchi et al. [7] | | Accuracy | 51.99 | 57.66 | - | 87.09 | 88.89 |
| | | | F−measure | 12.96 | 30.05 | - | 88.93 | 90.47 |
| | | | Success Rate | 0.9 | 3.44 | - | 99.93 | 100 |
| | Lin et al. [8] | | Accuracy | 50.82 | 51.33 | - | 53.63 | 79.44 |
| | | | F−measure | 3.42 | 6.25 | - | 18.03 | 65.19 |
| | | | Success Rate | 1.35 | 2.17 | - | 13.53 | 69.96 |

<div align="right">(<em>continued</em>)</div>

**Table 3.** (*continued*)

| QF1 | QF2 | | 55 | 65 | 75 | 85 | 95 |
|---|---|---|---|---|---|---|---|
| 85 | Proposed | Accuracy | **82.123** | **76.433** | **72.813** | **59.163** | **95.923** |
| | | F–measure | **70.733** | 59.503 | 61.263 | **38.313** | 86.203 |
| | | Success Rate | **90** | 40 | 40 | 0 | **100** |
| | Wang et al. [4] | Accuracy | 50.86 | 50.59 | 50.35 | - | 93.04 |
| | | F–measure | 43.72 | 55.05 | 64.02 | - | 93.85 |
| | | Success Rate | 58.82 | 77.28 | 91.63 | - | 99.55 |
| | Bianchi et al. [7] | Accuracy | 51.7 | 61.79 | 56.66 | - | 80.08 |
| | | F–measure | 26.2 | 67.93 | 27.55 | - | 84.08 |
| | | Success Rate | 4.33 | 80.42 | 1.87 | - | 99.55 |
| | Lin et al. [8] | Accuracy | 50.91 | 51.4 | 51.94 | - | 57.07 |
| | | F–measure | 3.84 | 6.54 | 10.14 | - | 22 |
| | | Success Rate | 1.72 | 2.62 | 4.48 | - | 23.84 |
| 95 | Proposed | Accuracy | **85.213** | **83.263** | **77.033** | **76.223** | **58.713** |
| | | F–measure | **76.213** | **72.033** | **65.853** | **69.583** | **37.133** |
| | | Success Rate | **100** | **100** | 70 | **85** | 0 |
| | Wang et al. [4] | Accuracy | 50.68 | 50.45 | 50.25 | 50.3 | - |
| | | F–measure | 45.74 | 55.25 | 65 | 57.5 | - |
| | | Success Rate | 60.91 | 77.58 | 93.27 | 82.29 | - |
| | Bianchi et al. [7] | Accuracy | 50.33 | 51.79 | 51.66 | 62.89 | - |
| | | F–measure | 24.15 | 26.3 | 34.99 | 56.25 | - |
| | | Success Rate | 4.04 | 3.44 | 5.68 | 36.4 | - |
| | Lin et al. [8] | Accuracy | 50.95 | 51.25 | 51.85 | 53 | - |
| | | F–measure | 3.93 | 6.32 | 10.71 | 19.67 | - |
| | | Success Rate | 1.42 | 2.84 | 4.86 | 11.21 | - |

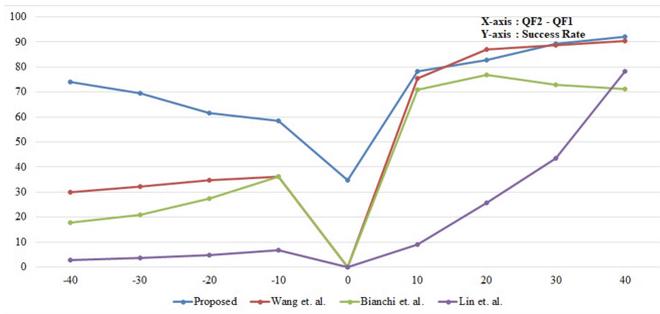

**Fig. 8.** Average Success Rate of Localization for varying $QF_2 - QF_1$ values

**Table 4.** Average accuracy, F–measure and Success Rate of Localization results, with varying Q2−Q1 for 10% forgery.

| Q2–Q1 | | Proposed | Wang et al. [4] | Bianchi et al. [7] | Lin et al. [8] |
|---|---|---|---|---|---|
| −40 | Accuracy | 83.573 | 37.77 | 73.19 | 88.04 |
| | F-measure | **72.033** | 12.48 | 10.32 | 1.48 |
| | Success Rate | **95** | 0 | 0 | 0 |
| −30 | Accuracy | 81.133 | 34.315 | 76.21 | 87.31 |
| | F-measure | **67.378** | 13.06 | 13.665 | 1.835 |
| | Success Rate | **60** | 0 | 0 | 0 |
| −20 | Accuracy | 75.98633 | 30.75 | 60.22 | 86.13667 |
| | F-measure | **60.26633** | 14.00333 | 13.98333 | 2.45 |
| | Success Rate | **36.66667** | 0 | 0 | 0 |
| −10 | Accuracy | 70.7855 | 30.93 | 80.1725 | 84.31 |
| | F-measure | **57.5155** | 14.385 | 27.8325 | 3.275 |
| | Success Rate | **30** | 0 | 3.025 | 0 |
| 0 | Accuracy | **58.473** | 0 | 0 | 0 |
| | F-measure | **34.571** | 0 | 0 | 0 |
| | Success Rate | 0 | 0 | 0 | 0 |
| 10 | Accuracy | 89.0905 | 91.36 | 71.3475 | 87.0775 |
| | F-measure | **77.47** | 66.02 | 50.99 | 2.13 |
| | Success Rate | **97.50** | 62.45 | 32.94 | 0.09 |
| 20 | Accuracy | 91.61 | 94.32 | 81.94 | 89.77 |
| | F-measure | **81.74** | 79.55 | 60.10 | 16.43 |
| | Success Rate | **100.00** | 78.23 | 40.23 | 15.89 |
| 30 | Accuracy | **94.83** | 94.66 | 76.05 | 91.39 |
| | F-measure | **88.11** | 82.59 | 52.87 | 31.99 |
| | Success Rate | **100.00** | 82.03 | 27.88 | 32.13 |
| 40 | Accuracy | 95.47 | 93.59 | 74.82 | 94.31 |
| | F-measure | **90.963** | 83.85 | 50.16 | 68.84 |
| | Success Rate | **100** | 83.56 | 20.4 | 70.1 |

## 4.3 Performance Evaluation and Comparison with State–of–the–Art

In this section, we present the performance evaluation results of the proposed method, as well as compare its performance with the state–of–the–art, in terms of all three parameters defined above (Sect. 4.2). We have compared the proposed method with three recent state–of–the–art techniques for JPEG forgery detection and localization, viz. the schemes of Wang et al. [4], Bianchi et al. [7] and Lin et al. [8].

Tables 1, 2, and 3, show the Forgery detection accuracy, F–measure and Success Rate of Localization results, of the proposed approach, along with the

**Table 5.** Average accuracy, F–measure and Success Rate of Localization results, with varying Q2–Q1 for 30% forgery.

| Q2–Q1 | | | Proposed | Wang et al. [4] | Bianchi et al. [7] | Lin et al. [8] |
|---|---|---|---|---|---|---|
| −40 | | Accuracy | **84.653** | 43.85 | 61.87 | 69.79 |
| | | F-measure | **74.133** | 31.84 | 19.18 | 3.3 |
| | | Success Rate | **100** | 0 | 0.22 | 0.22 |
| −30 | | Accuracy | **82.238** | 41.68 | 64.13 | 69.425 |
| | | F-measure | **69.538** | 33.975 | 22.67 | 4.065 |
| | | Success Rate | **80** | 0.15 | 0.185 | 0.185 |
| −20 | | Accuracy | **75.843** | 39.56667 | 58.24667 | 68.99667 |
| | | F-measure | **61.60967** | 36.41667 | 29.74667 | 5.44 |
| | | Success Rate | **45** | 0.146667 | 0.273333 | 0.223333 |
| −10 | | Accuracy | **70.113** | 39.0825 | 71.66 | 68.19 |
| | | F-measure | **58.348** | 37.825 | 40.185 | 7.5625 |
| | | Success Rate | **33.75** | 0.54 | 10.2375 | 0.335 |
| 0 | | Accuracy | **58.621** | 0 | 0 | 0 |
| | | F-measure | **35.007** | 0 | 0 | 0 |
| | | Success Rate | 0 | 0 | 0 | 0 |
| 10 | | Accuracy | 89.098 | 90.0475 | 80.565 | 69.8075 |
| | | F-measure | 78.01 | 79.08 | 76.67 | 10.24 |
| | | Success Rate | **98.75** | 78.10 | 72.26 | 3.79 |
| 20 | | Accuracy | 91.65 | 93.07 | 84.33 | 76.06 |
| | | F-measure | 82.64 | 90.24 | 81.14 | 29.29 |
| | | Success Rate | **100.00** | 91.98 | 84.90 | 26.61 |
| 30 | | Accuracy | **95.11** | 93.86 | 80.40 | 82.18 |
| | | F-measure | 89.33 | 91.33 | 77.48 | 48.86 |
| | | Success Rate | **100.00** | 93.87 | 75.71 | 49.89 |
| 40 | | Accuracy | **95.61** | 94.10 | 79.68 | 91.46 |
| | | F-measure | 91.943 | 92.59 | 76.31 | 81.96 |
| | | Success Rate | **100** | 94.92 | 74.96 | 85.43 |

methods proposed in [4,7,8]. Tables 1, 2, and 3, present the results for three different forgery sizes: 10%, 30%, 50% respectively, of the actual image, manually forged following the JPEG modification model described in Sect. 1 as well as Sect. 4.1.

As evident from Tables 1, 2, and 3, the diagonal entries, where $QF_1 = QF_2$, that is, the quality factors for the first and second compressions are same, the state–of–the–art methods fail; whereas, the proposed method is able to detect forgery with considerable efficiency. It is also evident that, in most of the cases the proposed method outperforms the others, specially for those cases where the first compression factor is greater than the second, i.e. $QF_1 > QF_2$.

**Table 6.** Average accuracy, F–measure and Success Rate of Localization results, with varying Q2–Q1 for 50% forgery.

| Q2–Q1 | | | Proposed | Wang et al. [4] | Bianchi et al. [7] | Lin et al. [8] |
|---|---|---|---|---|---|---|
| −40 | | Accuracy | **85.213** | 50.68 | 50.33 | 50.95 |
| | | F-measure | **76.213** | 45.74 | 24.15 | 3.93 |
| | | Success Rate | **100** | 60.91 | 4.04 | 1.42 |
| −30 | | Accuracy | **82.693** | 50.655 | 51.745 | 51.08 |
| | | F-measure | **71.383** | 49.485 | 26.25 | 5.08 |
| | | Success Rate | **95** | 68.2 | 3.885 | 2.28 |
| −20 | | Accuracy | **76.32967** | 50.51333 | 55.14667 | 51.35667 |
| | | F-measure | **63.18967** | 53.65333 | 38.62667 | 6.89 |
| | | Success Rate | 60 | 75.03667 | 29 | 2.943333 |
| −10 | | Accuracy | **70.8455** | 50.6675 | 60.2825 | 51.77 |
| | | F-measure | **59.868** | 56.6325 | 40.68 | 9.86 |
| | | Success Rate | 45 | 79.915 | 17.3025 | 4.745 |
| 0 | | Accuracy | **57.415** | 0 | 0 | 0 |
| | | F-measure | **34.747** | 0 | 0 | 0 |
| | | Success Rate | 0 | 0 | 0 | 0 |
| 10 | | Accuracy | **89.303** | 81.4525 | 84.0525 | 53.7175 |
| | | F-measure | 78.89 | 80.84 | 85.36 | 14.60 |
| | | Success Rate | **98.75** | 88.75 | 95.78 | 11.81 |
| 20 | | Accuracy | **91.78** | 89.55 | 87.31 | 62.20 |
| | | F-measure | 83.99 | 91.09 | 89.20 | 31.17 |
| | | Success Rate | **100.00** | 98.28 | 100.00 | 31.32 |
| 30 | | Accuracy | **95.58** | 91.50 | 85.67 | 71.35 |
| | | F-measure | 90.37 | 92.57 | 88.03 | 49.94 |
| | | Success Rate | **100.00** | 98.73 | 100.00 | 53.03 |
| 40 | | Accuracy | **96.07** | 93.40 | 84.32 | 88.49 |
| | | F-measure | 93.413 | 94.38 | 86.96 | 84.09 |
| | | Success Rate | **100** | 99.48 | 100 | 89.69 |

Existing literature proves that it is challenging to detect the tampered regions when $QF1 > QF2$, as the image behaves more like a single compressed image in this case. In terms of Accuracy, we find that the proposed method performs better than the state–of–the–art techniques, especially when $QF1 > QF2$. This is because, CNNs help to preserve the spatial structured features, and efficiently learn the statistical patterns of JPEG coefficient distribution, hence improving the detection accuracy.

In terms of F–measure, we can observe that the proposed method outperforms the state–of–the–art techniques in most cases even when the forgery size is 10%. It is evident from Tables 1, 2, and 3, that the F–measure results fall, as the forgery size increases.

Also, the proposed method achieves higher Success Rate of Localization, for forgery sizes of 10%, 30% and 50%, specially when $QF_1 > QF_2$. The thresholding on F–measure (in Sect. 4.2) indicates that we consider the successful cases, where 66.66% of the tampered region is correctly located.

**Performance with Varying Quality Factors.** In Tables 4, 5, and 6, we present the performance evaluation results of the proposed method, with different $QF_1$ and $QF_2$ values, specifically, for varying $QF_2 - QF_1$. The evaluation parameters used are the same as above. The results presented in Tables 4, 5, and 6, are the averages over different compression factors, producing a certain $QF_2 - QF_1$ value.

The results are also presented in form of 2D plots in Figs. 6, 7, and 8, for Accuracy, F–measure and Success Rate of Localization, respectively. The above plots are drawn, considering the average of 10%, 30% and 50% forgery sizes (of the entire image). The negative values on the left of the graphs, represent the cases where $QF_1 > QF2$ and the positive values on the right, represent the cases where $QF_1 < QF_2$.

It is evident from Figs. 6, 7, and 8, that for both all three cases, viz. $QF_1 > QF_2$, $QF_1 = QF_2$ and $QF_1 < QF_2$, the proposed JPEG forgery detection technique outperforms the other state–of–the–art methods. However, it performs best in case of $QF1 > QF2$, (left side of origin in the plots); whereas for rest two cases the superiority is marginal. In such cases, the accuracy can be improved further, by considering more (>7) number of blocks in the proposed method. (In this work we have considered only seven neighbouring blocks as described in Sect. 3.1.) But this also increases the training complexity parallely. Our finding is that, seven neighbouring blocks consideration, helps to attain a trade–off between performance efficiency and computational complexity.

## 5   Conclusion

In this paper, we propose a method to detect re–compression based JPEG image forgery, using deep neural network. We detect the presence of tampering in a JPEG, as well as locate the tampered region(s), based on a proposed CNN model, which is trained with the features of single compressed and double compressed image blocks. The inherent capability of automatic feature learning in deep CNNs, help us to achieve superior performance as compared to the state–of–the–art. Finally, the proposed CNN performs block–wise forgery prediction, for which we have considered nineteen DCT coefficients (second to twentieth in zig–zag order) from each block.

Our experimental results are encouraging and prove that the proposed techniques achieves considerably high forgery detection and localization efficiency, as compared to the state–of–the–art, especially when the first compression ration is greater than the second.

Future research in this direction would involve investigation of triple and higher degrees of JPEG compression based forgeries.

**Acknowledgement.** This work is partially funded by Board of Research in Nuclear Sciences (BRNS), Department of Atomic Energy (DAE), Govt. of India, Grant No. 34/20/22/2016-BRNS/34363, dated: 16/11/2016.